\documentclass[a4paper,fleqn,usenatbib]{mnras}
\usepackage{times,txfonts}
\usepackage[T1]{fontenc}
\usepackage{ae,aecompl}
\usepackage{color}

\usepackage{mnras_macros}
\usepackage{graphicx}	
\defcitealias{planck14}{Planck Collaboration 2014}

\newcommand{\RED}[1]{#1}

\title[The metal enrichment of passive galaxies]{The metal enrichment of passive galaxies in cosmological simulations of galaxy formation}
\author[T. Okamoto et al.]{Takashi Okamoto$^{1}$\thanks{E-mail:
  okamoto@astro1.sci.hokudai.ac.jp}, Masahiro Nagashima$^{2}$, 
  Cedric G. Lacey$^{3}$, and Carlos S. Frenk$^{3}$
\\
$^{1}$Department of Cosmosciences, Graduates School of Science, Hokkaido University, N10 W8, Kitaku, Sapporo, Hokkaido 060-0810, Japan\\
$^{2}$Faculty of Education, Bunkyo University, Koshigaya, Saitama 343-8511, Japan\\
$^{3}$Institute for Computational Cosmology, Department of Physics, Durham University, South Road, Durham, DH1 3LE}
\begin{document}

\date{Accepted . Received ; in original form }

\pagerange{\pageref{firstpage}--\pageref{lastpage}} \pubyear{2015}

\maketitle

\label{firstpage}

\begin{abstract}
  Massive early-type galaxies have higher metallicities and higher ratios of
  $\alpha$ elements to iron than their less massive counterparts. Reproducing
  these correlations has long been a problem for hierarchical galaxy formation
  theory, both in semi-analytic models and cosmological hydrodynamic simulations. 
  We show that a simulation
  in which gas cooling in massive dark haloes is quenched by radio-mode
  active galactic nuclei (AGNs) feedback naturally reproduces the observed trend
  between $\alpha$/Fe and the velocity dispersion of galaxies, $\sigma$.
  The quenching occurs earlier for more massive galaxies. Consequently, these
  galaxies complete their star formation before $\alpha$/Fe is diluted
  by the contribution from type Ia supernovae. For galaxies more
  massive than $\sim 10^{11}~M_\odot$ whose $\alpha$/Fe 
  correlates positively with stellar mass, we find an inversely correlated
  mass-metallicity relation. This is a common problem in simulations in which 
  star formation in massive galaxies is quenched either by quasar- or radio-mode AGN
  feedback. The early suppression of gas cooling in progenitors of
  massive galaxies prevents them from recapturing enriched gas ejected as winds.
  Simultaneously reproducing the [$\alpha$/Fe]-$\sigma$ relation and the
  mass-metallicity relation is, thus, difficult in the current
  framework of galaxy formation.
\end{abstract}


\section{Introduction}

The metal abundance of early-type galaxies  
provides strong constraints on the theory of their formation. 
Observational estimates indicate an increase of the overall metal
abundance with galaxy luminosity and velocity dispersion
\citep[e.g.][]{faber73, bender93}, and also an increase in the 
average ratio of $\alpha$ elements to iron 
abundance with luminosity and velocity dispersion 
\citep[e.g.][]{oconnel76, worthey92, jorgensen99, thomas10, 
johansson12, conroy13}.  


Theoretical modelling of chemical abundances in elliptical 
galaxies has been intensively investigated using the so-called 
monolithic collapse model, in which all of the stars 
are born in a single burst of short duration, which is usually 
assumed to be terminated by the ejection of the remaining gas by galactic 
winds \citep[e.g.][]{larson75, ay87}. 
Studies based on the monolithic collapse model
have shown that it is possible to explain the trends of both 
increasing total metallicity and increasing $\alpha$/Fe  
with galaxy mass \citep[e.g.][]{matteucci94, tgb99} if the 
star formation timescale of the initial burst is chosen 
to have a suitable dependence on galaxy mass.

Models and simulations based on the cold dark matter (CDM) cosmology 
have naturally reproduced the observed slope of 
the relation between stellar metallicity and velocity dispersion
(or stellar mass) by including strong feedback even in massive galaxies
\citep{kc98, nagashima05b, dave11, illustris}. 
On the other hand, the increase of $\alpha$/Fe with velocity 
dispersion has been a challenge for hierarchical galaxy formation 
models.  

Metal enrichment of elliptical galaxies in the context of hierarchical galaxy
formation has been intensively studied by utilising so-called semi-analytic
models.  \citet{thomas99} studied the chemical enrichment of elliptical
galaxies by considering  a closed box model for each galaxy using
pre-calculated star formation histories obtained by a semi-analytic model. He
claimed that an $\alpha$/Fe overabundant population cannot be obtained unless
the initial mass function (IMF) is significantly flattened with respect to the
Salpeter IMF \citep{salpeter} during starbursts.  \citet{nagashima05b}
investigated the chemical enrichment of ellipticals by using a semi-analytic
model in which they solved chemical evolution due to both type Ia and II
supernovae (SNe) self-consistently with galaxy merger trees.  Although they
reproduced the  observed $\alpha$ element abundance in ellipticals by assuming a
top-heavy IMF in starbursts, their models predict a decreasing $\alpha$/Fe
ratio with velocity dispersion, $\sigma$. 

The role of feedback from active galactic nuclei (AGNs) in chemical enrichment
of elliptical galaxies has also been investigated with semi-analytic
models.  Two distinct modes have been considered for AGN feedback: quasar
and radio modes. The former assumes that feedback energy and momentum are
proportional to mass accretion rate onto a black hole.  Feedback is thus
strongest during a rapid growing phase of a black hole \citep{dsh05}.
The latter has been introduced to prevent excessive growth of central galaxies
in massive haloes from cooling flows. Many semi-analytic models assume that the
radio mode is triggered by hot gas accretion \citep[e.g.][]{croton06, bower06};
this feedback is hence most efficient in massive haloes where the cooling time is
longer than the dynamical time. In numerical simulations, these two modes are
usually distinguished by the specific accretion rate onto a black hole.
When the accretion rate is small, typically less than 1\% of the Eddington
rate, feedback operates on the radio mode and directly heats the hot
halo gas \citep{sijacki07, onb08}.  
Quasar mode feedback, on the other hand, 
operates when the specific accretion rate is high and it heats the
interstellar medium around the black hole.

\citet{delucia06} demonstrated that, if radio-mode AGN feedback introduced by 
\citet{croton06} is included, massive galaxies have higher metal abundances, older 
luminosity-weighted ages, and shorter star formation time-scales.  
\citet{delucia12}, however, pointed out that semi-analytic models with
radio-mode AGN feedback predict decreasing stellar metallicity with
increasing stellar mass for massive galaxies, while they can explain old
stellar ages of massive galaxies. \citet{delucia06} and \citet{delucia12}
appear to reach the opposite conclusions about the effect of AGN feedback on the
metallicities in massive galaxies.  The more effective feedback in massive galaxies
employed by the models in \citet{delucia12} than in \citet{delucia06}
presumably causes this discrepancy.
\citet{pipino09} investigated the [$\alpha$/Fe]-$\sigma$ relation in ellipticals 
by including chemical enrichment both by SNe II and Ia in their semi-analytic model. 
They also considered chemical enrichment by low- and intermediate-mass stars and delay 
time distribution (DTD) of SNe Ia. 
Although they found a better but still marginal agreement with the observed [$\alpha$/Fe]-stellar 
mass relation by taking radio-mode AGN feedback into account, doing this erases a positive 
correlation between stellar metallicity and velocity dispersion. 

\citet{calura09} implemented chemical enrichment model on top of star formation 
histories obtained from a semi-analytic model as done in \citet{thomas99}. 
They summed star formation histories over all progenitors of a galaxy in question to 
compute chemical evolution. 
They assumed that the IMF becomes top-heavy in systems with star formation rates (SFRs) 
greater than $100~M_\odot~\mathrm{yr}^{-1}$. 
This assumption, together with quasar-mode AGN feedback and interaction-triggered starbursts, 
yields a better agreement with the [$\alpha$/Fe]-$\sigma$ and mass-metallicity relations 
than the case where the Salpeter IMF is assumed for all star formation activities. 

\citet{calura11} employed the same model as \citet{calura09}, but computed chemical evolution 
using individual star formation histories of all progenitors of a galaxy considered.
They claimed that inclusion of quasar-mode AGN feedback that quenches late star formation 
in large haloes and starbursts triggered by galaxy harassment that boosts star formation 
activity at high redshift are crucial and the variable IMF used in \citet{calura09} 
in not necessary to explain the observed [$\alpha$/Fe]-$\sigma$ relation, 
although they did not show the mass-metallicity relation predicted by their model. 

\citet{arrigoni10} also investigated the [$\alpha$/Fe]-mass and mass-metallicity
relations by using a semi-analytic model that includes both quasar- and radio-modes 
AGN feedback and a detailed chemical evolution model in which they relax the instantaneous 
recycling approximation. 
They claimed that a combination of a shallower slope for the IMF and a low SN Ia rate 
(i.e. a low binary fraction) is required to obtain a good agreement with observations. 
\citet{yates13} implemented three different SN Ia DTDs  
in their semi-analytic model. They found that a combination of a power-law DTD and  
metal-rich winds that drive light $\alpha$ elements directly out of galaxies 
gives the best agreement with a wide range of observational data including 
the [$\alpha$/Fe]-mass and mass-metallicity relations. 
On the other hand, \citet{gargiulo15} employed a variable IMF whose slope becomes 
shallower (more top-heavy) for a higher SFR of a galaxy. 
They tuned their model to reproduce the observed [$\alpha$/Fe]-galaxy mass relation
and they claimed that the SFR dependent IMF is key to reproducing 
the [$\alpha$/Fe]-galaxy mass relation and a short star formation time-scale due 
to radio-mode AGN feedback alone cannot explain the slope of the relation. 

The numerous semi-analytic studies have shown that it is difficult 
to simultaneously reproduce the observed [$\alpha$/Fe]-$\sigma$ and mass-metallicity relations.
Different authors propose different solutions, most of which involve some form of 
AGN feedback, variable IMFs, or a combination of these two. 

This problem has not been extensively studied by using cosmological 
hydrodynamic simulations, which, in principle, have advantages 
in dealing with chemical enrichment over semi-analytic models, 
that is, (i) chemical enrichment is solved locally in hydrodynamic 
simulations rather than galaxy-wide and (ii) the instantaneous 
recycling approximation is easily relaxed by using star particles, 
each of which represents a simple stellar population, and 
by solving timed release of mass, metals, and energy from 
each star particle \citep[e.g.][]{oka05}. 

Some cosmological hydrodynamic simulations also found 
a decreasing $\alpha$/Fe ratio with velocity dispersion 
\citep{romeovelona13, taylor15}. 
\citet{taylor15} further showed that quasar-mode AGN feedback can increase 
$\alpha$/Fe in massive galaxies without changing the IMF. 
\RED{\citet{segers16b} also showed that [$\alpha$/Fe] increases with stellar 
mass for massive galaxies whose stellar masses are more massive than $10^{10.5}~M_\odot$ 
when AGN feedback is included, while it is roughly constant without AGN feedback.}  

Recent advances in computer simulations have enabled us to 
form reasonably realistic galaxy populations in a representative 
cosmological volume 
\citep[e.g.][]{okamoto14, illustris, eagle}. 
In this paper we investigate the chemical abundances of 
passive galaxies formed in cosmological hydrodynamic simulations 
with the physics described by \citet[][hereafter OSY14]{okamoto14}, 
which reproduces key observed properties of galaxy populations, 
such as the galaxy stellar mass function, star formation rates, 
and gas phase metallicities of star-forming galaxies, 
for $z \le 4$. 
We discuss the role of each physical process in determining 
the chemical properties of the passive galaxies and reveal 
successes and limitations of our galaxy formation model. 

This paper is organised as follows. In Section 2, we briefly 
describe our simulations. 
We present our results in Section 3.  
The results are summarised and discussed in Section 4.

\section[]{Simulations and sample of galaxies}

We study the galaxies formed in cosmological hydrodynamic 
simulations in the $\Lambda$-dominated CDM ($\Lambda$CDM) 
cosmology with the following parameters \citepalias{planck14}: 
$\Omega_0 = 0.318$, $\Omega_\Lambda = 0.682$, 
$\Omega_\mathrm{b} = 0.049$, $\sigma_8 = 0.835$, 
$n_\mathrm{s} = 0.962$, and dimensionless Hubble parameter,  
$h = 0.67$. 
We use a cosmological periodic box of side length $80~h^{-1}$~Mpc
for the reference simulation. 
The simulation code is based on an early version of Tree-PM smoothed 
particle hydrodynamics (SPH) code {\scriptsize GADGET-3}. 
We employ $512^3$ dark matter particles and the same number of 
SPH particles. 
Each gas particle initially has a mass of 
$m_\mathrm{SPH}^\mathrm{orig} = 5.2 \times 10^7 h^{-1}~M_\odot$.  
The gravitational softening length is set to $2.1~h^{-1}$~kpc 
in physical units both for gas and star particles after z = 3.  

The baryonic processes and the numerical resolution are exactly 
the same as in OSY14.  
We briefly describe the implementation of the feedback
and chemical enrichment processes. 
Star formation takes place in the dense cold gas, and the star 
formation timescale is assumed to be proportional to the local 
dynamical time as 
\begin{equation}
  \dot{\rho}_* = c_* \frac{\rho_\mathrm{gas}}{t_\mathrm{dyn}}, 
\end{equation}
where $c_*$ and $t_\mathrm{dyn}$ are, respectively, 
the dimensionless star formation efficiency parameter and 
the local dynamical time. 
We employ $c_* = 0.01$ in our simulations.  
If the mass of an SPH particle, $m_\mathrm{SPH}$, is 
more massive\footnote{In our simulations, the mass of an SPH 
particle can change by star formation and feedback.} 
than $1.5 m_*$, where $m_* = 0.5   m_\mathrm{SPH}^\mathrm{orig}$, 
we compute the probability ${\cal P}_\mathrm{spawn}$ for an SPH particle with which 
it spawns a new star particle of mass, $m_*$, during a timestep, $\Delta t$, as 
\begin{equation}
  {\cal P}_\mathrm{spawn} = \frac{m_\mathrm{SPH}}{m_*} 
  \left[1 - \exp\left(-\frac{c_* \Delta t}{t_\mathrm{dyn}}\right)\right]. 
\end{equation}
When $m_\mathrm{SPH}$ is smaller than $1.5 m_*$, we compute the probability, 
${\cal P}_\mathrm{conv}$, with which 
the SPH particle is converted into a star particle of mass, $m_\mathrm{SPH}$, 
during $\Delta t$ as  
\begin{equation}
  {\cal P}_\mathrm{conv} = 1 - \exp\left(-\frac{c_* \Delta t}{t_\mathrm{dyn}}\right),  
\end{equation}
in order to avoid the SPH particle mass becomes too small. 

A newly born star particle inherits smoothed (SPH-kernel averaged) 
chemical abundances of its parent gas particle as in 
\citet{oka05, ofjt10}. 
Each star particle represents a simple stellar population (SSP) 
and its evolution is modelled as in \citet{okamoto13}. 
We employ the Chabrier IMF \citep{chabrier03} and use 
metallicity-dependent stellar lifetimes and chemical yields 
\citep{pcb98, marigo01}. 
We calculate type II and Ia SN rates based 
on \citet{pcb98}  (see also \citealt{nagashima+05a}). 
We track 9 elements (H, He, C, N, O, Ne, Mg, Si, and Fe) and 
we take S and Ca to proportional to Si \citep{wie09b}. 
To normalise chemical abundances of simulated galaxies by the solar values, 
we employ solar abundances compiled by \citet{asplund09}, and thus $Z_\odot = 0.0142$. 
On the other hand, we do not make any corrections to published [$\alpha$/Fe] or 
metallicity values\footnote{Observational estimates presented in this paper 
all assume the old canonical solar metallicity, $Z_\odot = 0.02$},  
since we assume that observations measure relative abundances; 
applying the solar abundances by \citet{asplund09} for the observational estimates, 
however, does not affect our conclusions at all. 

Stellar feedback is modelled as winds. 
Massive stars emit a large amount of ultraviolet photons in their early 
phases and then die as SNe II.   
We follow the luminosity evolution of each star particle 
and calculate how much photon energy and momentum are given to surrounding gas particles 
during a given timestep. 
The feedback due to radiation pressure is modelled as 
``momentum-driven winds'' \citep{od06}, and we assume that the initial 
wind speed is scaled with the local dark matter velocity dispersion, $\sigma_\mathrm{DM}$. 
The wind mass loading is, hence, proportional to $\sigma_\mathrm{DM}^{-1}$ 
for a given momentum input. 
We also calculate, for each star particle, the number of stars that die as SNe II during 
a timestep.
The energy released by SNe II is distributed to gas particles around the star particles.
Feedback by SNe II is treated as ``energy-driven winds''  
\citep{ofjt10}. 
Again, the initial wind speed is scaled with 
$\sigma_\mathrm{DM}$ and therefore the wind mass loading is proportional 
to $\sigma_\mathrm{DM}^{-2}$. 
For gas particles that receive radiation pressure or SN energy, we 
attach probabilities with which they are added to winds. 
We then stochastically select wind particles and increase their velocities.

It is now widely accepted that stellar feedback alone 
cannot explain the bright-end of the galaxy luminosity 
function and hence feedback processes that operate preferentially 
in massive galaxies are required \citep[e.g.][]{ben03}. 
We introduce a simple, phenomenological analogue of the {\it radio-mode} 
AGN feedback \citep[e.g.][]{croton06, bower06, onb08}. 
We simply assume that radiative cooling is suppressed for gas 
particles around which one-dimensional dark matter velocity dispersion 
is larger than $\sigma_\mathrm{th}(z)$, where $z$ is redshift.  
We parameterise the functional form of $\sigma_\mathrm{th}$  as 
$\sigma_\mathrm{th}(z) = \sigma_0 (1 + z)^\alpha$. 
In order to reduce the cooling rate in massive haloes, as the radio-mode 
feedback does, we modify the cooling function, $\Lambda(T, Z)$, as 
\begin{eqnarray}
  \Lambda(T, Z, \sigma)=\left\{ \begin{array}{ll}
      \Lambda(T, Z) & (\sigma < \sigma_{\rm th}) \\
      \Lambda(T, Z) \exp\left(-\frac{\sigma - \sigma_{\rm th}}{\beta \sigma_{\rm th}}\right) & (otherwise), \\
  \end{array} \right.
  \label{eq:agn}
\end{eqnarray}
where $T$ and $Z$ are respectively the gas temperature and the gas metallicity, 
and the parameter $\beta$ specifies how steeply the cooling is suppressed 
above $\sigma_\mathrm{th}$. We adopt $\sigma_0 = 50$~km~s$^{-1}$, 
$\alpha = 0.75$, and $\beta = 0.3$ in our reference simulation. 
This suppression of gas cooling is quite effective at low redshift.  
The cooling rate is halved in a halo with $\sigma \simeq 60$~km~s$^{-1}$ 
at $z = 0$ from the original cooling rate. 
We note that comparable results to this set of parameters can be obtained 
when we assume $\Lambda(T, Z, \sigma) = \Lambda(T, Z)$ for $\sigma < \sigma_\mathrm{th}$ 
and $\Lambda(T, Z, \sigma) = 0$ for $\sigma > \sigma_\mathrm{th}$ with $\sigma_0 = 100$~km~s$^{-1}$ 
and $\alpha = 0.75$ (OSY14).  
The cooling is thus suppressed even in modest groups at $z = 0$; the effect becomes almost 
negligible at high redshift ($z \gtrsim 2$) owing to the redshift dependence in $\sigma_\mathrm{th}(z)$. 
We refer to this suppression of the cooling in massive haloes as 
``AGN feedback'' in this paper. 
We also show the results for a simulation without AGN feedback, 
which is the simulation labelled `SN+RP' in OSY14, 
for highlighting the effect of this process. 
The size of the simulation box, ($40~h^{-1}$~Mpc)$^3$, is 
smaller than that of the reference simulation. 

To identify virialized dark matter haloes, we run the friends-of-friends
(FoF) group finder \citep{defw85} with a linking length of 0.2 in units of 
the mean dark matter particle separation. 
Baryonic particles near dark matter particles that compose an FoF group 
are also regarded as members of the halo. 
We then use the {\scriptsize SUBFIND} algorithm \citep{spr01} 
to identify gravitationally bound sub-groups of particles 
within each FoF halo. 
We regard a sub-group that consists of at least 32 particles  
and contains at least 10 star particles as a ``galaxy".  
We then define radius of a galaxy in order to exclude 
diffusely distributing stellar components as done in  
\citet{vogelsberger13}. 
We first calculate the stellar half-mass radius, $r_\mathrm{h}$, for each 
galaxy and define galaxy radius as $2 r_\mathrm{h}$. 
This definition of galaxy radius has almost no effect on the properties 
of low mass galaxies, but it excludes some of the intracluster light 
for massive galaxies. 
\citet{vogelsberger13} show that this definition gives very similar results 
to more elaborate methods of excluding intracluster light. 

We remove galaxies that do not contain more than 10 star 
particles within their $r$-band half-light radii from our sample. 
We compute chemical abundances of a galaxies as luminosity- 
or mass-weighted average of chemical abundances of the star particles 
within a specified radius (e.g. half-light radius and galaxy radius). 
We calculate the three-dimensional velocity dispersion of the star particles 
within the $r$-band half-light radius and then convert it into the one-dimensional 
velocity dispersion by multiplying a factor, $1/\sqrt{3}$. 
We have confirmed that we obtain quantitatively similar results when we use 
projected half-light radius and line-of-sight velocity dispersion. 

In addition, we analyse public data of the Illustris simulation  
(Illustris-1) \citep{illustris_data} and the EAGLE simulation 
(RefL0100N1504) \citep{eagle_data} for comparison.  
We show the results whenever corresponding data are available. 
The details of the Illustris simulation are found in a series 
of papers \citep{illustris_intro, illustris}. 
Radius, chemical abundances, and velocity dispersion of Illustris galaxies 
are defined in the same way as in OSY14.

The EAGLE simulation is fully described in \citet{eagle} and \citet{eagle_rob}.  
In the EAGLE simulation, galaxies are identified by using the FoF and 
{\scriptsize SUBFIND} algorithms as in OSY14 and Illustris.  
Stars gravitationally bound to a sub-group and within a spherical 
30 kpc aperture are used to define stellar mass, luminosity, and velocity 
dispersion of the galaxy \citep{eagle}. 
The chemical abundances of individual elements of a galaxy in the public
database are, however, calculated as the mass-weighted average of chemical
abundances of stars bound to a subhalo. We will use these mass-weighted
chemical abundances to compute abundance ratios of EAGLE galaxies. 

There are some important differences between the reference, Illustris, and EAGLE 
simulations. 
While stellar feedback in the reference and Illustris simulations 
are implemented as winds, feedback energy is deposited locally as thermal energy in 
the EAGLE simulation. 
For AGN feedback, the reference simulation only has the radio-mode feedback. 
AGN feedback in the Illustris simulation has two modes: quasar and radio modes, while 
the EAGLE simulation only has the quasar-mode feedback. 

\subsection{Identification of passive galaxies}

\begin{figure}
 \begin{center}
  \includegraphics[width=\linewidth]{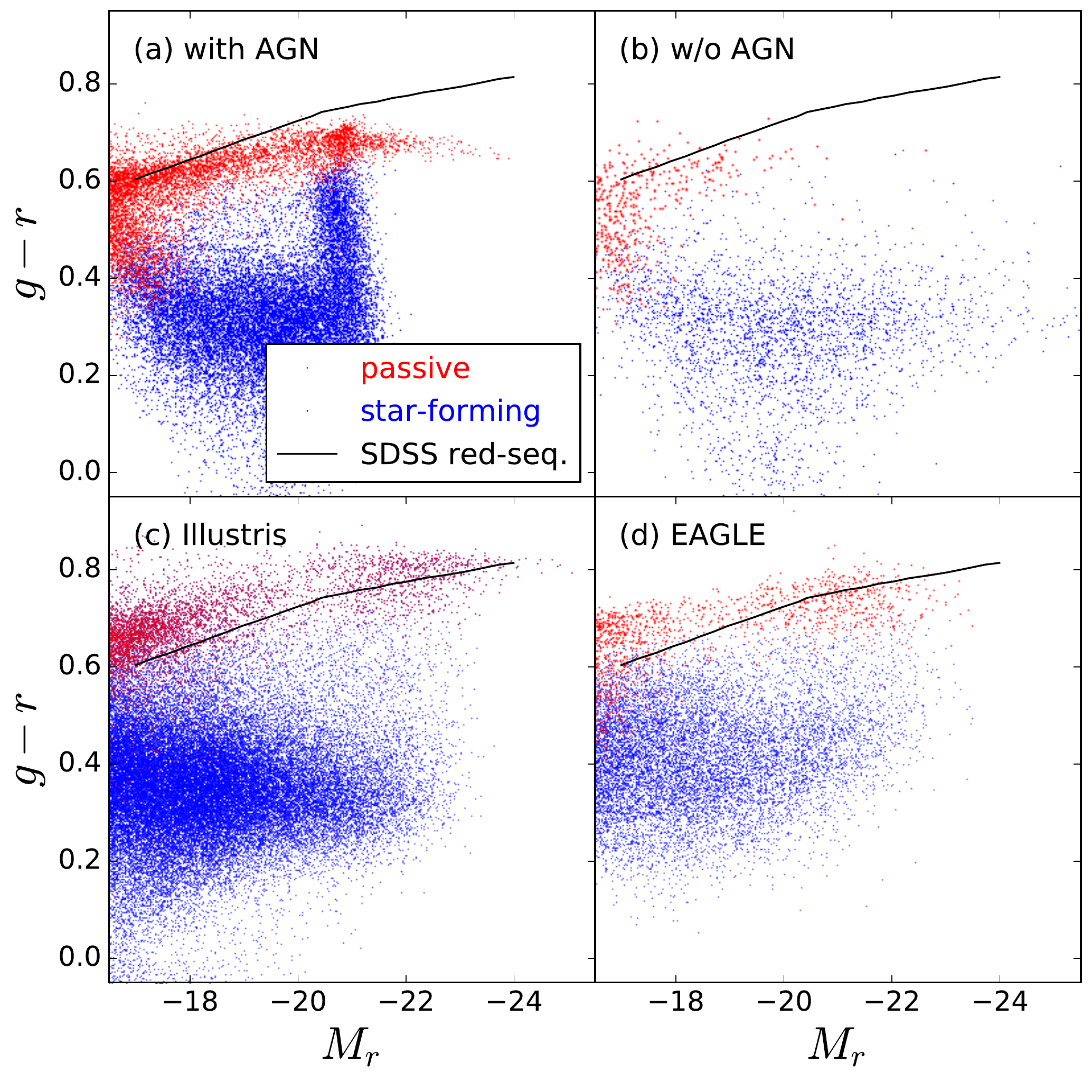} 
 \end{center}
 \caption{ 
   Colour-magnitude diagrams for the simulated galaxies. 
   We plot $g - r$ colours for (a) the reference simulation, 
   (b) the simulation without AGN feedback, (c) the Illustris 
   simulation, and (d) the EAGLE simulation.  
   We indicate passive galaxies and star-forming galaxies by red and blue 
   dots, respectively. 
   We also show the colour-magnitude relation 
   for SDSS red sequence measured by \citet{skelton09} by solid lines.
}
\label{fig:cmr}
\end{figure}

Morphological classification of simulated galaxies is difficult 
and could be affected by the limited numerical resolution. 
We hence focus on the chemical abundances of passive galaxies. 
We select galaxies whose specific star formation rates 
are smaller than $10^{-11}$~yr$^{-1}$ as the passive galaxies. 
In Fig.~\ref{fig:cmr}, we show the colour-magnitude diagram of 
simulated galaxies in the reference simulation, the 
simulation without AGN feedback, the Illustris simulation, 
and the EAGLE simulation. 
For these plots, we use all the star particles belonging to a galaxy 
to compute luminosities and colours of the simulated galaxy  
and compare them with the colour-magnitude relation  for 
SDSS red sequence measured by \citet{skelton09}. 

There is clearly a red sequence and a blue cloud in the reference, Illustris, and EAGLE 
simulations,  
while there are only a few red bright galaxies in the simulation without AGN feedback. 
In the reference simulation, there is a bridge between the 
red sequence and the blue cloud at $M_r \simeq -21$, where star-forming 
galaxies are being quenched by radio-mode feedback. 
We would be able to erase this feature by introducing quasar mode 
feedback that more quickly terminates star formation
as in the Illustris and EAGLE simulations. 

The colours of the passive galaxies in the reference simulation are 
somewhat bluer than the SDSS red sequence. 
We find that metallicities of passive galaxies would need 
to be doubled in order to agree with observations. 
The chemical yields assumed here, however, can explain the chemical 
abundances of the Local Group satellites \citep{ofjt10} and 
Milky Way-mass galaxies \citep{no06}. 
Passive galaxies in the Illustris simulation are slightly redder than the 
SDSS red sequence, reflecting higher metallicities than those in the reference simulation 
as we will show later. 
The EAGLE simulation reproduce the red sequence reasonably well, although 
the faint-end slope is slightly shallower than the data 
\RED{(see also \citet{eagle_color} for EAGLE galaxies)}. 
Galaxies with specific star formation rates smaller than $10^{-11}$~yr$^{-1}$ 
lie on the red sequence in all simulations and this simple criterion successfully 
selects passive galaxies.

\section{Results}

\subsection{The $\alpha$-to-Fe ratios of simulated galaxies} 

\begin{figure}
 \begin{center}
  \includegraphics[width=\linewidth]{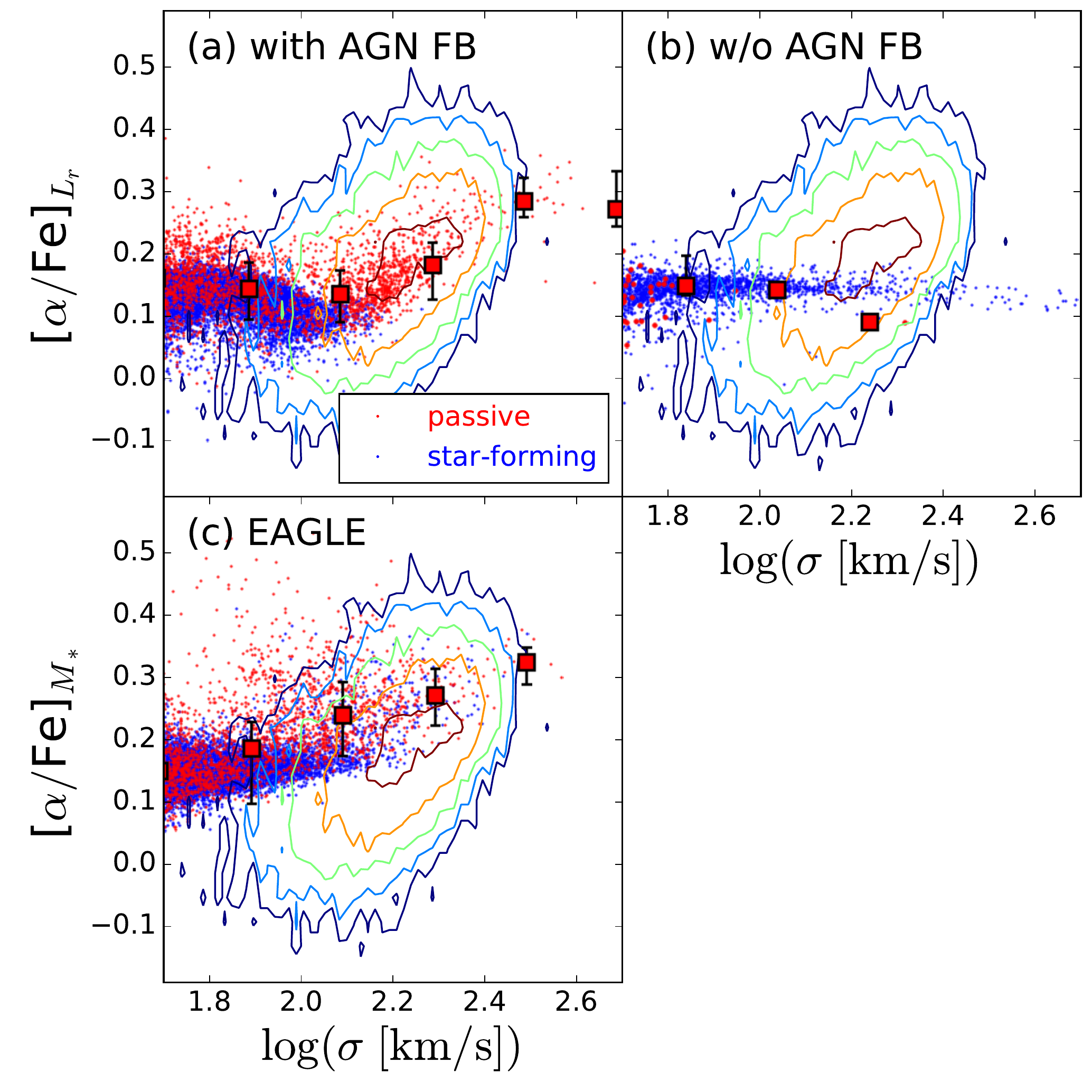} 
 \end{center}
 \caption{Luminosity-weighted [$\alpha$/Fe] versus velocity dispersion of 
   galaxies in (a) the reference simulation and  (b) the simulation 
   without AGN feedback. 
   We also show the mass-weighted [$\alpha$/Fe] of Eagle galaxies in 
   panel (c).  
   Passive and star-forming galaxies are, respectively, indicated by 
   the red and blue dots. 
   The median [$\alpha$/Fe] in a velocity bin and 16th to 84th percentile 
   distribution for the passive galaxies are indicated by a red square 
   with an error bar. 
   Observational estimates for  
   SDSS passive early-type galaxies \citep{thomas10} are shown by the 
   contours.} 
\label{fig:afe}
\end{figure}

In Fig.~\ref{fig:afe}, we plot the [$\alpha$/Fe]-$\sigma$ relations for  
galaxies in the reference simulation, the simulation without AGN feedback, 
and the EAGLE simulation. 
We do not show Illustris galaxies because the public data of the Illustris 
simulation do not include abundances of $\alpha$ elements.
We also show the observational estimates by \citet{thomas10} for  
the MOSES sample \citep{schawinski09, thomas10}. 
We only use passive galaxies in MOSES to compare 
with the passive galaxies in the simulations. 
For the simulations by OSY14, we show luminosity weighted [$\alpha$/Fe], 
while, for the EAGLE simulation, we show mass-weighted [$\alpha$/Fe]. 

From convergence studies in OSY14, we find that ages and metallicities 
of the simulated galaxies in the reference simulation and that without AGN feedback 
are reliable down to $M_* \sim 10^{9.5} M_\odot$; 
this stellar mass corresponds to the velocity dispersion, 
$\sigma \sim 10^{1.6}~\mathrm{km~s}^{-1}$. 
We hence expect that the [$\alpha$/Fe] relations shown in 
Fig.~\ref{fig:afe} are not affected by the numerical resolution. 
We have also confirmed that a result obtained by a simulation with higher 
resolution (but with a smaller box size) agrees with that in 
Fig.~\ref{fig:afe}.

The reference simulation reproduces the observational trend for high $\sigma$
galaxies well;  [$\alpha$/Fe] of the passive galaxies increases with the
velocity dispersion for $\sigma \gtrsim 100$~km~s$^{-1}$.  Below $\sigma \sim
100$~km~s$^{-1}$, [$\alpha$/Fe] becomes nearly constant, while the
observational trend continues at least down to $\sigma \sim
10^{1.7}~\mathrm{km~s}^{-1}$ \citep[e.g.][]{spolaor10, johansson12}. The star
forming galaxies, in the reference simulation, exist below $\sigma \sim
10^{2.1}$~km~s$^{-1}$ and have slightly lower [$\alpha$/Fe] than the passive
galaxies.  In the simulation without AGN feedback, [$\alpha$/Fe] does not
depend on the velocity dispersion, and the value is consistent with that for
$\sigma \lesssim 100$~km~s$^{-1}$ in the reference simulation.
\RED{\citet{segers16b} showed similar results with the EAGLE simulation, 
that is, [$\alpha$/Fe] increases with stellar mass for galaxies with stellar 
masses of $M_* > 10^{10.5}~M_\odot$ when AGN feedback is included, while it is roughly 
constant without AGN feedback.}

\RED{As expected from the results by \citet{segers16b}}, 
the EAGLE simulation also shows an increasing 
[$\alpha$/Fe] for passive 
galaxies with velocity dispersion, while the relation extends to lower 
velocity dispersion and its slope is somewhat shallower than in the reference 
simulation. 
It is interesting that almost all simulated galaxies (both passive and star-forming) have 
[$\alpha$/Fe]$ > 0$. 
This is partly because low mass simulated galaxies are too old compared with 
their observed counterparts (OSY14). 
Top-light IMFs for low mass galaxies and/or a delayed distribution of SNe Ia with an 
increased normalisation compared with 
the DTD currently assumed would also decrease [$\alpha$/Fe] of low mass galaxies 
without changing [$\alpha$/Fe] of massive galaxies. 

\begin{figure}
 \begin{center}
  \includegraphics[width=\linewidth]{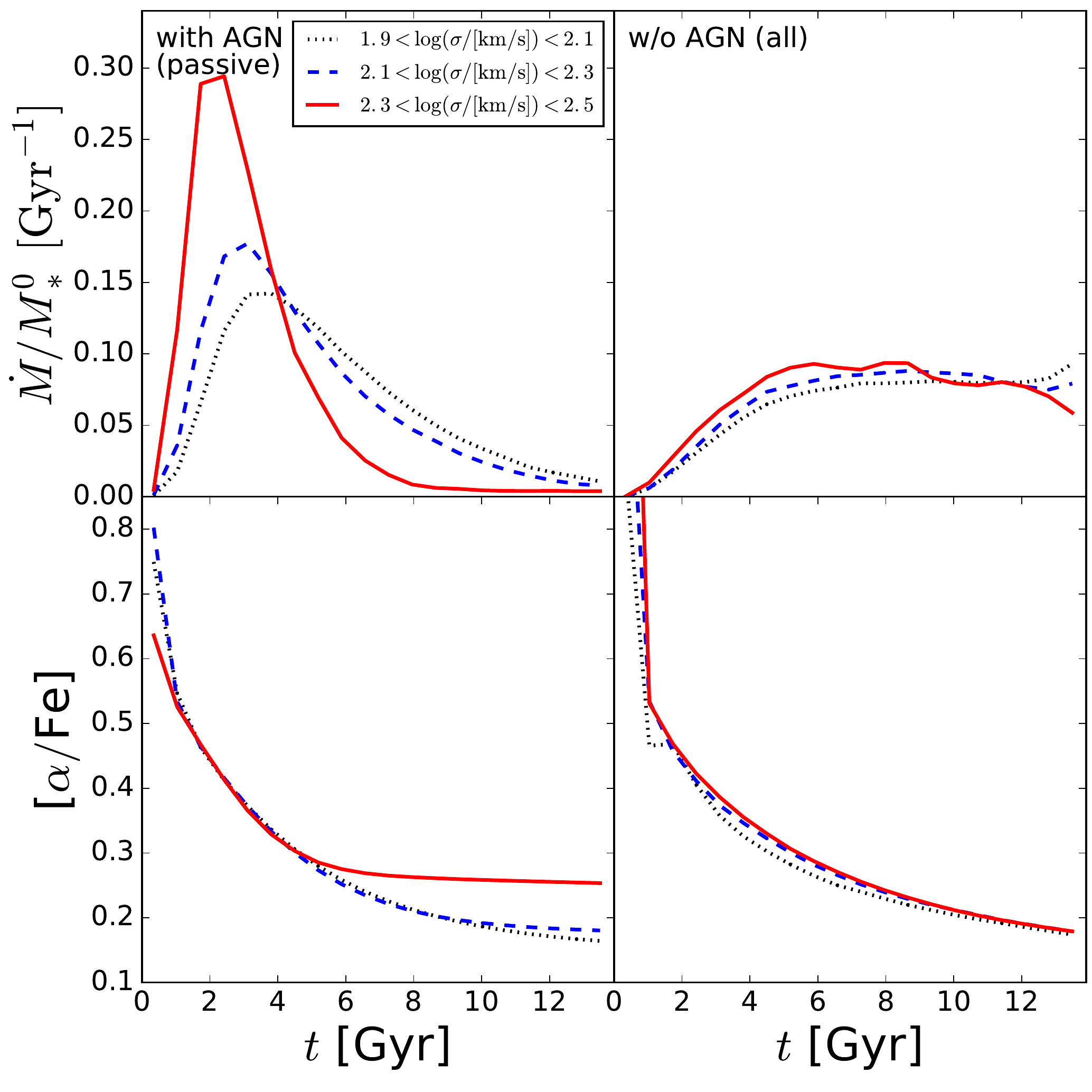} 
 \end{center}
 \caption{ 
   {\it Upper panels}: Mean specific star formation rates (normalised by the stellar 
   mass at $z = 0$) as a function of cosmic time in three velocity bins at $z = 0$: 
   $10^{1.9} < \sigma/\mathrm{[km~s}^{-1}\mathrm{]} < 10^{2.1}$ 
   (black dotted), 
   $10^{2.1} < \sigma/\mathrm{[km~s}^{-1}\mathrm{]} < 10^{2.3}$ 
   (blue dashed), and 
   $10^{2.3} < \sigma/\mathrm{[km~s}^{-1}\mathrm{]} < 10^{2.5}$ (red solid). 
   The formation histories of the stars within the half light radius at 
   $z = 0$ are shown. 
   Left and right panels, respectively, show the reference simulation and 
   the simulation without AGN feedback. We only use passive galaxies for the reference 
   simulation, while all the galaxies in a velocity bin are used for the simulation 
   without AGN feedback, which are mostly star-forming.
   {\it Lower panels}: Same as the upper panel but for [$\alpha$/Fe].
   We here show the mass-weighted [$\alpha$/Fe] of the stars that have formed by 
   time $t$. 
}
\label{fig:sfrafe}
\end{figure}

To understand the origin of the [$\alpha$/Fe]-$\sigma$ relation for $\sigma
\gtrsim 100$~km~s$^{-1}$ in the reference simulation, we investigate the star
formation histories and the evolution of the stellar [$\alpha$/Fe] for galaxies
having different velocity dispersion at $z = 0$.  We show the normalised
formation histories and the time evolution of [$\alpha$/Fe] of stars within
the half right radii at $z = 0$, dividing galaxies into three velocity bins in
Fig.~\ref{fig:sfrafe}.  In the reference simulation, galaxies with high
$\sigma$ have a peak in the star formation history at an earlier epoch and a
shorter star formation timescale than those with lower $\sigma$.  This is a
direct consequence of our AGN feedback model that suppresses gas cooling
earlier in haloes with higher velocity dispersion.  As shown in
Fig.~\ref{fig:sfrafe}, the longer star formation timescale for lower $\sigma$
galaxies reduces [$\alpha$/Fe] with time \RED{as discussed in \citet{segers16b}}.  
On the other hand, in the
simulation without AGN feedback, the normalised star formation histories are 
almost identical in all the velocity bins.  
As a result, the evolution of
[$\alpha$/Fe] becomes independent of the velocity dispersions at $z =
0$.

\subsection{The stellar metallicities of simulated galaxies}

\begin{figure}
 \begin{center}
  \includegraphics[width=\linewidth]{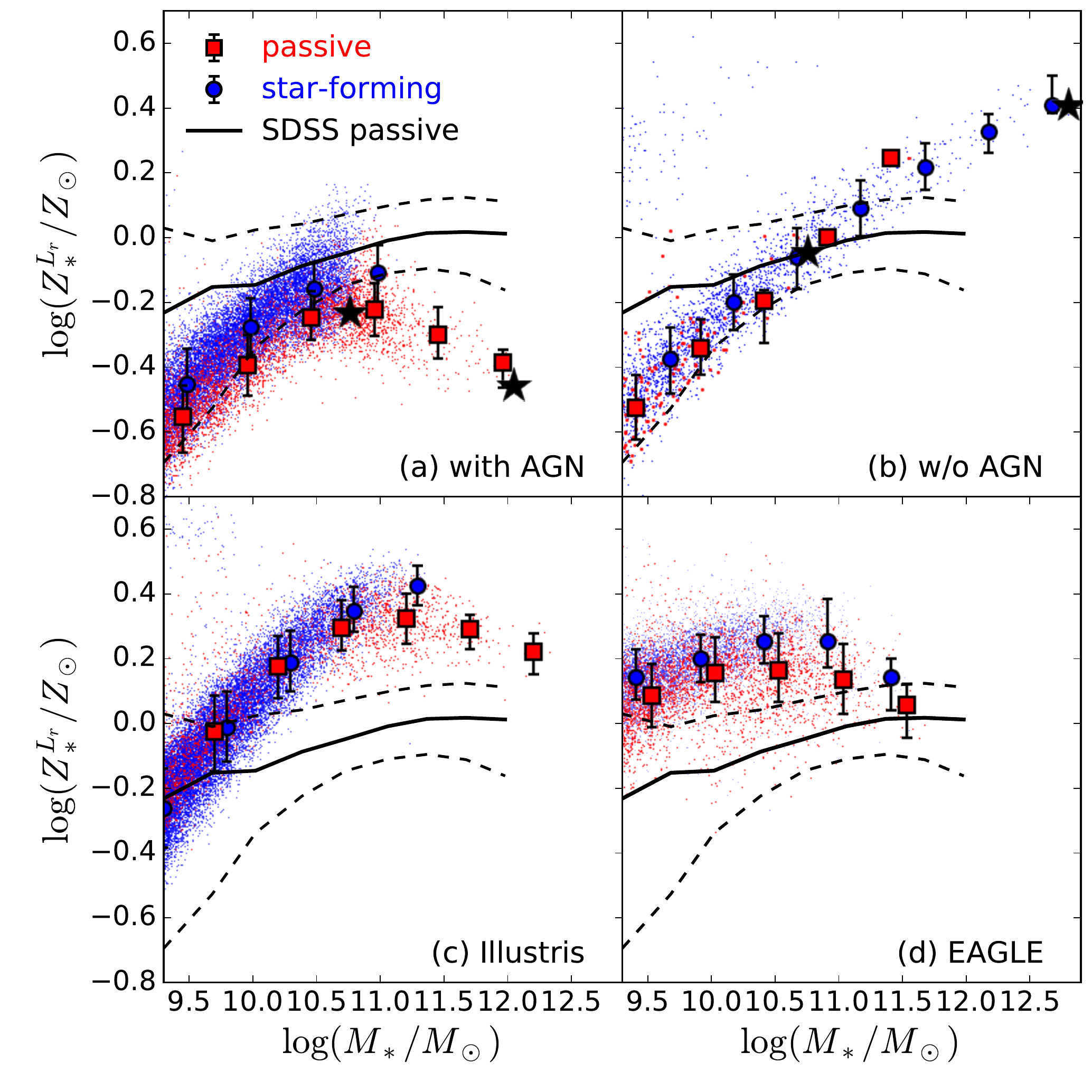} 
 \end{center}
 \caption{    
   Stellar mass-stellar metallicity relation for the simulated galaxies. 
   We show $r$-band luminosity-weighted metallicities for (a)
   the reference simulation, (b) the simulation without AGN feedback, and (c)
   the Illustris simulation.  Passive and star-forming galaxies are,
   respectively, indicated by the red and blue dots. The squares show the
   median metallicities in the mass bins for the passive galaxies and the
   circles for the star-forming galaxies.  The observational estimates for SDSS
   passive galaxies are shown by the black lines; the solid line indicates the
   median of the distribution and the dotted lines the 16th and 84th
   percentiles, where we apply the aperture correction as described in
   Appendix~\ref{sec:sdss}.  The black stars in the panel (a) indicate two
   passive central galaxies that we investigate their enrichment histories
   later in Fig.~\ref{fig:enrichment}.  
   In the panel (b), we also highlight two (star-forming) galaxies 
   by black stars, for which we show their enrichment histories in Fig.~\ref{fig:enrichment}, too. 
}
\label{fig:mzr}
\end{figure}

The reference simulation reproduces well the gas phase metallicity of the
star-forming galaxies (OSY14). 
We now examine the stellar metallicities of the simulated galaxies. 
In the left panels of Fig.~\ref{fig:mzr}, we show luminosity-weighted 
stellar mass-stellar metallicity relations (MZRs) 
of galaxies in the reference simulation, 
in the simulation without AGN feedback, in the Illustris simulation, 
and in the EAGLE simulation.  

For the simulations by OSY14 and the Illustris simulation, we calculate
the $r$-band luminosity-weighted average of metallicities of stars bound to a
subhalo and within a galaxy radius defined as twice the stellar half-mass radius.
For the EAGLE simulation, we use all stars bound to a subhalo to estimate the
metallicity instead of applying a 30 kpc aperture because, at $M_* \gtrsim 10^{11}~M_\odot$, 
the galaxy radius in the reference and Illustris simulations is larger than 30 kpc and
monotonically increases with stellar mass. 
We also show the MZR of SDSS passive galaxies, 
where we correct the observed metallicity for each galaxy from 
an aperture value to a global one (see Appendix~\ref{sec:sdss}).  

In the reference simulation, the metallicities of passive galaxies 
are lower than the observational estimate. 
For low mass galaxies with $M_* \lesssim 10^{10.5} M_\odot$, 
the MZR is steeper than the observational estimate, whereas 
more massive galaxies, $M_* \gtrsim 10^{10.5} M_\odot$, 
show a trend of decreasing metallicity with increasing stellar mass. 
A similar trend is found for Illustris galaxies, although 
they on average have higher metallicities than the galaxies in 
the reference simulation. 
The higher metallicities of the Illustris galaxies compared with those 
in the reference simulation 
are explained by the fact that, in Illustris, the metallicity of a newly-created 
wind particle is chosen to be 40\% of the metallicity of the ambient interstellar 
medium, while, in OSY14, all the metal mass attached to a gas particle is ejected 
when the particle is launched as a wind.
The mass range where metallicity negatively correlates with stellar 
mass is the mass range where AGN feedback operates.  
In fact, the stellar metallicities in the simulation without AGN 
feedback monotonically increase with the stellar mass. 

We find that galaxies in the EAGLE simulation have much flatter MZR than 
those in the reference and Illustris simulations. 
The slope is more consistent with the observational estimate for the 
passive galaxies. 
The massive EAGLE galaxies, however, show a weak trend of decreasing metallicity 
with increasing stellar mass as those in the reference and the Illustris 
simulations. 
\RED{This trend is consistent with the finding of \citet{segers16a} using the EAGLE simulation. 
They showed that the fraction of stellar mass contributed 
by recycled stellar-ejecta, in the EAGLE simulation, increases with stellar mass for less 
massive galaxies and this fraction turns over at $M_* \sim 10^{10.5}~M_\odot$ owing to 
AGN feedback.} 
This feature seems to be common in simulations that invoke AGN feedback. 
We, however, note that the metallicities of passive galaxies monotonically
increase with stellar mass when we employ a 30 kpc aperture for the EAGLE
simulation. The same applies to the Illustris and reference simulations. We
must be careful about metallicity gradients within galaxies when comparing
simulated and observed metallicities.

A similar problem has been reported using semi-analytic models. 
\citet{pipino09} successfully produce an increasing [$\alpha$/Fe] 
with galaxy stellar mass by invoking radio-mode AGN feedback. They, however, 
find that the AGN feedback erases a positive correlation between stellar 
metallicity and velocity dispersion. 
Similarly, \citet{delucia12} find that feedback that makes ages of massive 
elliptical galaxies as old as observational estimates results in too low 
metallicities of the massive elliptical galaxies. 
\citet{mccarthy10} show that hydrodynamic simulations including quasar mode AGN 
feedback predict central group galaxies that have too low metallicity 
compared with observational data, while, in simulations without 
AGN feedback, the metallicities of central galaxies become higher than 
observed. 
We will investigate the origin of this problem later.

For less massive galaxies with $M_* \lesssim 10^{11}~M_\odot$,
the luminosity-weighted metallicities of passive galaxies in the reference 
simulation are, on average, slightly lower than those of star-forming galaxies. 
The EAGLE simulation shows a similar trend. In the Illustris simulation, passive
galaxies have comparable metallicities to star-forming galaxies. 
Although measuring stellar metallicities of star-forming galaxies is difficult, 
tentative observational estimates suggest that the luminosity-weighted metallicities of 
passive galaxies are significantly higher than those of star-forming 
galaxies \citep{peng15}. 
If passive galaxies do in fact have much higher metallicities 
than star-forming galaxies, we would need to consider a process that 
preferentially enriches passive galaxies. 

\begin{figure*}
 \begin{center}
  \includegraphics[width=\linewidth]{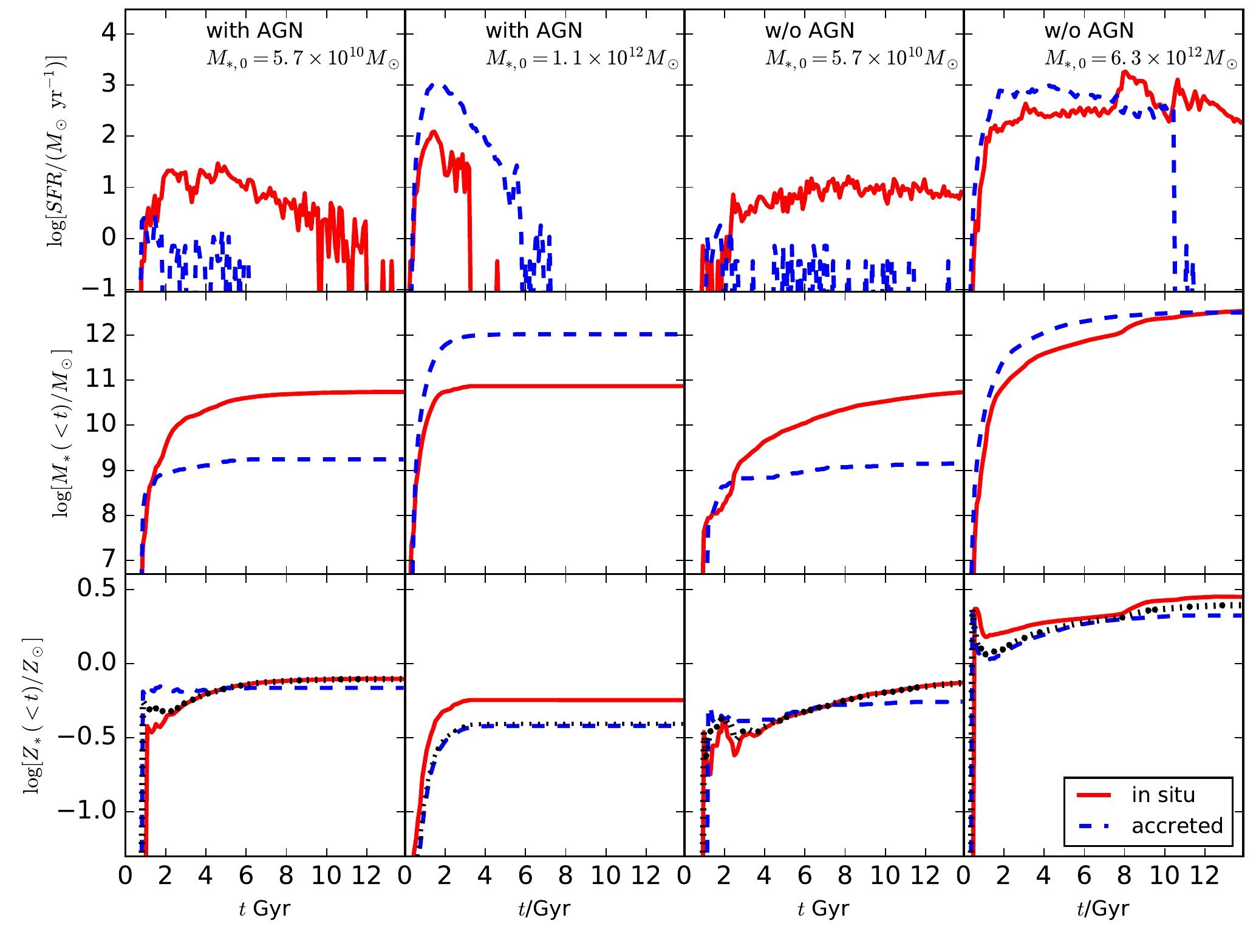} 
 \end{center}
 \caption{Formation and enrichment histories of selected central galaxies.  
  In the 1st and 2nd columns, we show two central passive galaxies in the reference simulation
  with stellar masses of $5.7\times 10^7 M_\odot$ and $1.1
  \times 10^{12} M_\odot$, respectively.  
  In the 3rd and 4th columns, we
  show galaxies in the simulation without AGN feedback.  The less massive
  galaxy has similar stellar mass to that in the reference simulation ($M_*
  \simeq 5.7\times 10^7 M_\odot$).  The galaxy shown in the 4th column is the
  most massive galaxy ($M_* \simeq 6.3 \times 10^{12}~M_\odot$) in this
  simulation. 
  These galaxies are indicated by the stars in the panels (a) and (b) of Fig.~\ref{fig:mzr}.  
  From top to bottom, we show star formation rates, masses of
  stars that have formed by time $t$, and mass-weighted average metallicities
  of stars that have formed by time $t$. Formation and enrichment histories of
  stars that formed in the main progenitor of each galaxy ({\it in situ}\/) are
  indicated by the red solid lines, while those of stars born in sub clumps are
  shown by the blue dashed lines. 
  In the bottom panels, we show total ({\it in situ} + accreted) metallicities 
  by the black dotted lines.
}
\label{fig:enrichment}
\end{figure*}

To understand the reason why simulations with AGN feedback generally yield
inversely correlated MZR, we investigate formation and enrichment histories of
galaxies. From the reference simulation, we select a central passive
galaxy at the knee of the MZR and the most massive galaxy. 
The masses and metallicities of these galaxies at $z = 0$ are
$(M_{*, 0}, Z^{L_r}_{*, 0}) \simeq (5.7 \times 10^{10}~M_\odot,~0.59~Z_\odot)$ and $(1.1
\times 10^{12}~M_\odot,~0.35~Z_\odot)$.  
For comparison, 
we pick two central star-forming galaxies, $(M_{*, 0}, Z^{L_r}_{*, 0}) 
\simeq (5.7 \times 10^{10}~M_\odot,~0.9~Z_\odot)$ and $(6.3
\times 10^{12}~M_\odot,~2.6~Z_\odot)$, from the simulation without 
AGN feedback; the latter is the most massive galaxy in the simulation. 
While the most massive galaxy in the simulation without AGN feedback 
is more massive than that in the reference simulation, 
its halo mass, $M_\mathrm{vir} \simeq 2 \times 10^{14}~M_\odot$, is smaller 
than that in the reference simulation, $M_\mathrm{vir} \simeq 4 \times 10^{14}~M_\odot$. 

In Fig.~\ref{fig:enrichment}, we show formation and enrichment histories of
{\it in situ} and {\it accreted} stars within these galaxies.  {\it In situ}
stars are the stars born in the main progenitor of each galaxy and  {\it
accreted} stars are those born in subclumps. 

The main progenitor of the less massive galaxy in the reference simulation
continues its star formation down to low redshift. On the other hand, AGN
feedback quenches the formation of {\it in situ} stars in the most massive
galaxy at early times ($t \sim 3$~Gyr). Consequently, this galaxy is dominated
by accreted stars at $z = 0$. The mass fractions of {\it in situ} stars in the
less massive and the most massive galaxies are 0.97 and 0.066 at $z = 0$,
respectively.  The metallicities of the less massive and most massive galaxies
thus, respectively, reflect the metallicities of {\it in situ} and accreted stars.

The main progenitor of the most massive galaxy always has a higher
velocity dispersion than that of the less massive galaxy and hence it has 
fewer outflows per star formed than the main progenitor of the less massive
galaxy\footnote{Wind mass loading scales with $\sigma^{-2}$ for SN-driven winds
and $\sigma^{-1}$ for radiation-driven winds in our simulations.}.  
We, however, find that {\it in situ} stars in the most massive
galaxy have lower metallicity than in the less massive galaxy. The
earlier suppression of gas cooling by AGN feedback in larger haloes prevents
massive galaxies from recapturing enriched outflows and renders their
metallicities lower than in their less massive counterparts as discussed by
\citet{delucia06}. 

In the simulation without AGN feedback, the formation of {\it in situ} stars 
of both galaxies continues to $z = 0$. Even without AGN feedback,
the fraction of {\it in situ} stars decreases with increasing galaxy mass as
indicated by \citet{naab14}. The metallicities of both {\it in situ} and
accreted stars of the most massive galaxy in this simulation are much higher
than those in the reference simulation. 

The low metallicity of accreted stars
of the most massive galaxy in the reference simulation suggests that subclumps 
in high-density regions also undergo AGN feedback. In fact, the SFR of 
the accreted stars in the most massive galaxy of the reference simulation shows 
a clear sign of early (but later than {\it in situ} stars) quenching.


\section{Summary and discussion}

We have analysed the chemical abundances of simulated passive 
galaxies. The reference simulation has previously been shown 
to reproduce many observed properties of galaxies at $z \le 4$, 
such as the stellar mass function, 
star formation rates, ages, and gas phase metallicities 
of star-forming galaxies (OSY14). 
We have also analysed publicly available data of two independent 
cosmological simulations: the Illustris \citep{illustris} and 
the EAGLE \citep{eagle} simulations. 
We define passive galaxies as those whose specific star formation 
rates are smaller than $10^{-11}$~yr$^{-1}$. 
This definition selects galaxies that lie on a red sequence.

We find that the [$\alpha$/Fe]-$\sigma$ relation of 
passive galaxies in the reference simulation is consistent 
with observations, that is, 
[$\alpha$/Fe] increases with velocity dispersion 
for $\sigma > 100$~km~s$^{-1}$. 
Passive galaxies in the EAGLE simulation also show 
the same trend. 
In the reference simulation, star formation in galaxies with high 
velocity dispersion today is quenched earlier by AGN feedback than 
in galaxies with lower velocity dispersion.   
Consequently, galaxies with higher velocity dispersions have shorter 
star formation timescales and older stellar ages than those with 
lower velocity dispersion. 
This process explains the trend in the [$\alpha$/Fe]-$\sigma$ relation 
at $\sigma \gtrsim 100$~km~s$^{-1}$. 
For less massive galaxies with $\sigma \lesssim 100$~km~s$^{-1}$, which are
not affected by AGN feedback, [$\alpha$/Fe] is nearly constant; 
this behaviour might be inconsistent with the observations which suggest 
that the [$\alpha$/Fe]-$\sigma$ trend exists at least for $\sigma \gtrsim 10^{1.7}$~km~s$^{-1}$  
\citep{spolaor10, johansson12}. 

Earlier quenching of star formation in more massive galaxies by AGN feedback, 
however, also decreases the stellar metallicity for increasing 
stellar mass. 
In spite of different implementations of physical processes, 
three independent simulations that invoke AGN feedback produce 
inversely correlated stellar mass to metallicity relations for 
massive galaxies, which contradict  
observational estimates. 

We investigate the enrichment and formation histories of passive galaxies with
different masses in the reference simulation.  We find that metallicities
of both {\it in situ} and accreted stars in  massive galaxies are lower than
those in less massive galaxies because of early quenching of gas cooling in
massive galaxies, which prevents enriched winds from falling back later. The
shorter formation time-scale of the accreted stars in the massive galaxy
compared with its counterpart in the simulation without AGN feedback suggests
that not only the main progenitor but also other progenitors are subject to AGN
feedback.  Simultaneously explaining the increasing $\alpha$-to-Fe ratio
with velocity dispersion (or stellar mass) and increasing stellar metallicity
with stellar mass is therefore fundamentally difficult within the current framework 
of galaxy formation. One possible solution
would be the introduction of a variable IMF that becomes more top-heavy for a
higher SFR as suggested by \citet{calura09} and \citet{gargiulo15} and previously 
suggested on independent grounds by \citet{baugh05} and \citet{lacey16}.  

Luminosity-weighted metallicities of passive galaxies in 
the reference simulation are, on average, lower than those 
of star-forming galaxies. 
Passive galaxies in the EAGLE simulation also have slightly lower 
metallicities than star-forming galaxies for a given stellar mass. 
In the Illustris simulation, passive galaxies have similar luminosity-weighted 
metallicities to star-forming galaxies. 
Tentative observational estimates suggest that the luminosity-weighted metallicities of passive 
galaxies are significantly higher than those of star-forming galaxies for a given stellar 
mass \citep{peng15}.  
None of the three simulations can produce this metallicity gap. 
A process that preferentially enriches passive galaxies would explain this 
observational result. 
More accurate measurements of stellar metallicity will thus  
place strong constraints on galaxy formation theory. 

\section*{Acknowledgments}

We thank Masato Onodera for his helpful comments on the 
observational data. 
Numerical simulations were carried out with Cray XC30 in CfCA 
at NAOJ. TO acknowledges the financial support of Japan Society
for the Promotion of Science (JSPS) Grant-in-Aid for Young Scientists
(B: 224740112) and MEXT KAKENHI Grant (16H01085). 
MN acknowledges the financial support of JSPS Grant-in-Aid (B: 25287041). 
CSF acknowledges ERC Advanced Investigator grant COSMIWAY.
This work was supported in part by STFC Consolidated grant ST/L00075X/1.
We also thank the Illustris collaboration for making data 
publicly available. 
We acknowledge the Virgo Consortium for making their simulation data 
available. The EAGLE simulations were performed using the DiRAC-2 
facility at Durham, managed by the ICC, and the PRACE facility 
Curie based in France at TGCC, CEA, Bruy\`eres-le-Ch\'atel.

\bibliography{okamoto_ref}

\appendix

%
%

\section{Total metallicities of SDSS passive galaxies} \label{sec:sdss}

\begin{figure}
 \begin{center}
  \includegraphics[width=\linewidth]{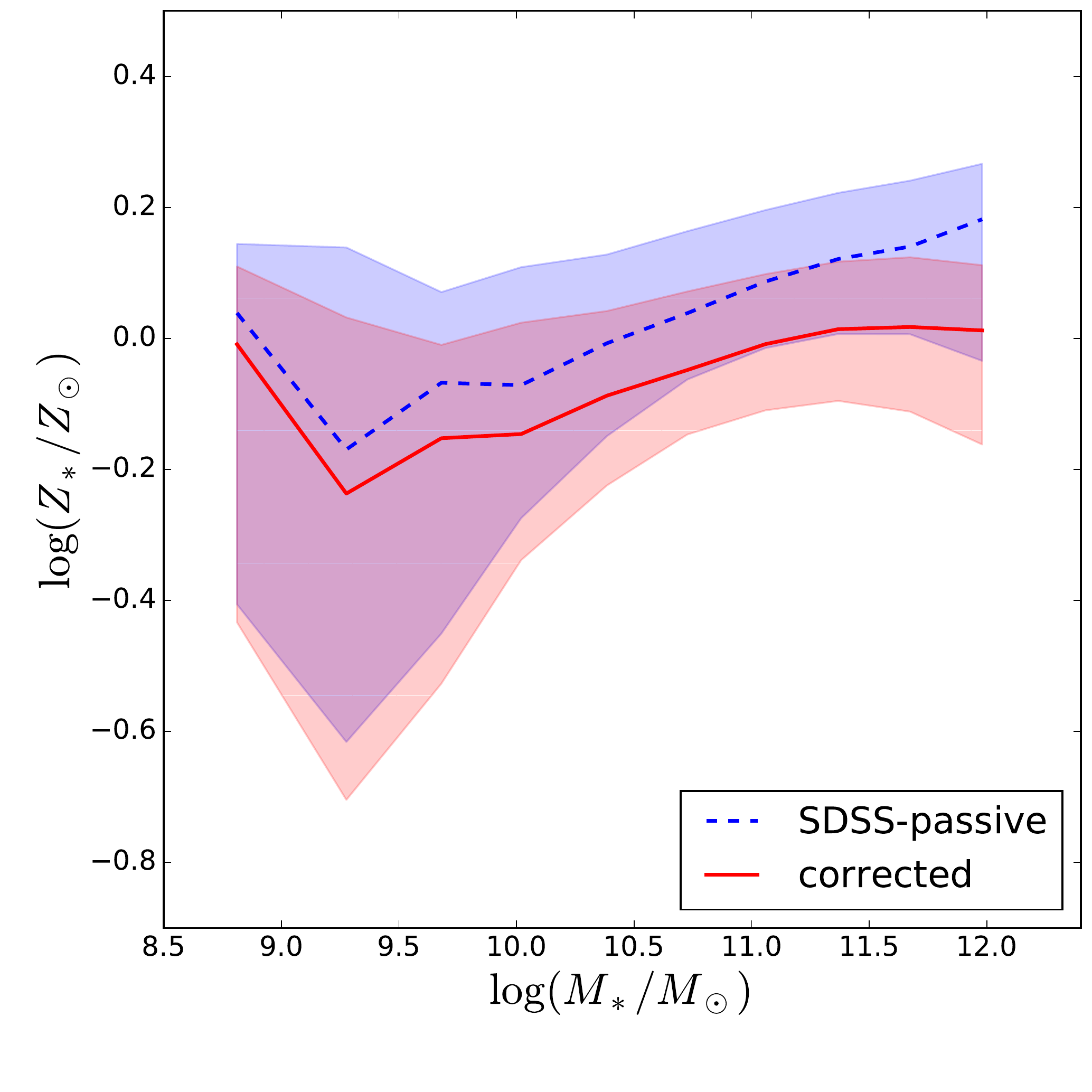} 
 \end{center}
 \caption{ 
  Mass-metallicity relation for SDSS-passive galaxies. 
  Stellar metallicities and masses are taken from \citet{gallazzi05}. 
  The median MZR obtained from the aperture values is shown by the blue 
  dashed line. 
  We correct the observed metallicity for each galaxy from an aperture value to a global one 
  by using redshift of each galaxy and assuming a fixed metallicity gradient. 
  The median MZR obtained this way is indicated by the red solid line. 
  The shaded regions represent the 16th to 84th distributions. 
  Note that we assume $Z_\odot = 0.02$ as in \citet{gallazzi05}. 
  }
\label{fig:sdss}
\end{figure}

To obtain an observational estimate for stellar metallicities of passive galaxies, 
we take metallicity and stellar mass estimates of the SDSS DR4 galaxies from \citet{gallazzi05}. 
We only use galaxies with $S/N > 20$. 
We select galaxies whose specific SFRs are smaller than $10^{-11}$~yr$^{-1}$ as candidates of 
passive galaxies by taking SFR estimates from \citet{brinchmann04}.  
We then remove galaxies that are classified as star-forming or composite from the passive galaxy 
candidates. 
The median MZR of the passive galaxies obtained this way is indicated by the dashed 
line in Fig.~\ref{fig:sdss}. 
This relation is consistent with the MZR of early-type galaxies in \cite{gallazzi06} 
and that of passive galaxies in \cite{peng15}. 

Early-type galaxies are known to have radial metallicity gradients \citep[e.g.][]{wu05, rawle10}.  
Since the original spectra were measured in 1.5 arcsec radius fibre aperture in SDSS, 
we correct the observed metallicities from aperture values to mean global values assuming 
a uniform metallicity gradient, $\mathrm{d}\log Z/\mathrm{d} \log r = -0.15$, based on 
\citet{rawle10} and by using the redshift of each galaxy.  
We here assume that the light profile in each galaxy follows an $r^{1/4}$ raw and 
the median size-mass relation for early-type galaxies by \citet{shen03}. 
The corrected median MZR is shown by the solid line in Fig.~\ref{fig:sdss}. 

\bsp

\label{lastpage}

\end{document}